\documentclass[final,12pt]{clear2026}

\newcommand{\Prob}[0]{\mathbb{P}}

\newcommand{\indep}{\perp \!\!\! \perp}

\title[Information-Theoretic Signatures of Causality]{Information-theoretic signatures of causality in Bayesian networks and hypergraphs}
\usepackage{times}
\usepackage[singlelinecheck=false,margin=0pt]{caption}
\usepackage{bm}
\usepackage{enumitem}

\clearauthor{
 \Name{Sung En Chiang} \Email{sung.chiang24@imperial.ac.uk}\\
 \addr Department of Mathematics, Imperial College London
 \AND
 \Name{Zhaolu Liu} \Email{zhaolu.liu16@imperial.ac.uk}\\
 \addr Department of Mathematics, Imperial College London
 \AND
 \Name{Robert L. Peach} \Email{peach\_r@ukw.de}\\
 \addr Department of Neurology, University Hospital W\"urzburg \\
 \addr Department of Brain Sciences, Imperial College London
 \AND
 \Name{Mauricio Barahona} \Email{m.barahona@imperial.ac.uk}\\
 \addr Department of Mathematics, Imperial College London
}

\begin{document}

\maketitle

\begin{abstract}
Analyzing causality in multivariate systems involves establishing how information is generated, distributed and combined. Traditional causal discovery frameworks are capable of multivariate reasoning but their intrinsic pairwise graph topology restricts them to do so only indirectly by integrating multivariate information across pairwise edges. Higher-order information theory provides direct tools that can explicitly model higher-order interactions. In particular, Partial Information Decomposition (PID) allows the decomposition of the information that a set of sources provides about a target into redundant, unique, and synergistic components. Yet the mathematical connection between such higher-order information-theoretic measures and causal structure remains undeveloped.
Here we establish the first theoretical correspondence between PID components and causal structure in both Bayesian networks and hypergraphs.
We first show that in Bayesian networks unique information precisely characterizes direct causal neighbors, while synergy identifies collider relationships. This establishes a localist causal discovery paradigm in which the structure surrounding each variable can be recovered from its immediate informational footprint, eliminating the need for global search over graph space. Extending these results to more expressive causal representation, we prove that PID signatures in Bayesian hypergraphs differentiate parents, children, co-heads, and co-tails, revealing a novel collider effect unique to multi-tail hyperedges. 
Our results position PID as a rigorous, model-agnostic foundation for inferring both pairwise and higher-order causal structure, and introduce a fundamentally local information–theoretic viewpoint on causal discovery.
\end{abstract}


\section{Introduction}

An important task in statistical and machine learning research is to understand how complex multivariate systems are structured: which variables influence which; how information is shared between them; in what ways their interactions extend beyond simple correlations. 
Classical structural models like Bayesian networks encode some of those relationships through directed edges, and are interpreted via conditional factorization and independence~\citep{10.5555/1642718, spirtes2000causation}. 
While these edge-based models in principle are able to capture complex higher-order interactions, inferring them requires aggregating information across the global structure since graphs intrinsically encode pairwise relations in local topology. Such higher-order interactions are crucial in many real-world domains, including neural, biological and social systems~\citep{luppi2022synergistic,schneidman2003synergy,iacopini2019simplicial} systems, where higher-order structures (e.g., hypergraphs or simplicial complexes) offer more accurate representations of the system and the data~\citep{torres2021and, bick2023higher}. Owing to this, in parallel,  a rich body of work on higher-order information-theoretic measures from Shannon information theory~\cite{shannon1948mathematical} has produced a range of tools for quantifying higher-order interactions~\citep{liu2024information,rosas2019quantifying, bell2003co}, yet their relationship to causality has remained conceptually underdeveloped. 

Among information-theoretic higher-order measures, Partial Information Decomposition (PID) offers a principled framework for analyzing the multivariate information that a set of sources conveys about a target, separating it into redundant, unique, and synergistic components~\citep{williams2010nonnegative},  
and thereby capturing the higher-order information-theoretic behavior that traditional graphical models are unable to represent.
However, despite extensive use across biomedicine~\citep{varley2023partial, luppi2024synergistic, cang2020inferring} and machine learning~\citep{liang2023quantifying, liang2024foundations, makkeh2025general}, PID has so far been primarily used as a descriptive metric, and lacks a structural interpretation that relates its components to graphical roles. Recent frameworks that introduce PID to causal analysis either rely on known temporal orderings to measure past-to-future information flow~\cite{mediano2025toward,martinez2024decomposing} or are constrained by additive noise assumptions and single-tail restrictions~\citep{enouen2025higher}.

To address these limitations, here we provide the first systematic correspondence between PID components and structural roles in both pairwise and higher-order graph structures, yielding a localist interpretation of causality. Standard causal discovery frameworks~\citep{kalisch2007estimating, 10.5555/1642718, spirtes2000causation} often execute local conditional independence tests, yet their underlying mechanics are fundamentally global. Because they are designed to construct a consistent network-wide equivalence class, a single edge orientation inevitably triggers widespread constraint propagation across the graph to prevent cycles and false v-structures. Similarly, score-based methods may utilize mathematically decomposable metrics, but these scores ultimately serve merely as heuristics to guide an algorithmic search such as the hill-climbing algorithm~\cite{tsamardinos2003time,tsamardinos2006max} under global structural constraints. In contrast, our PID-based localist perspective of causality focuses on one region of the system at a time: for any variable of interest, we study its immediate neighborhood by computing the relevant PID components to characterize the structural roles of the variables surrounding it in the (hyper)graph. 

We start by developing this mathematical correspondence for Bayesian networks~\citep{bishop2007}. In particular, we find that unique information in PID corresponds to direct causal neighbors, and the edge directions are identified through the synergistic patterns, yielding a two-step procedure. 
We then extend these concepts to Bayesian hypergraphs~\citep{javidian2020hypergraph}, a generalization of Bayesian networks that incorporates mixtures of directed and undirected relations and can naturally encode higher-order dependence. 
We establish an analogous correspondence between hypergraph structure and PID patterns: unique information characterizes membership in the tail or head of a hyperedge (or symmetric co-head relations) and synergy isolates co-tail structure, thus revealing the higher-order collider effect unique to hypergraphs. Together, these signatures specify the full informational footprint of a hyperedge. Furthermore, since a particular PID pattern may be compatible with multiple combinations of hyperedges, we developed a procedure for identifying hyperedges with local information patterns by selecting the largest (maximal) hyperedge consistent with the pattern. 

Our contributions are: (i) We enrich the multivariate reasoning with fine-grained interaction decomposition by establishing a formal link between multivariate Shannon information theory and causal structures by mapping PID components to causal roles in graphical models.
(ii) We show that this mapping yields a localist information-theoretic characterization of causal neighborhoods in Bayesian networks.
(iii) We extend this framework to Bayesian hypergraphs, a higher-order representation that explicitly encode multivariate information.
(iv) We provide procedures that can be implemented to provide systematic causal discovery in both Bayesian networks and hypergraphs using this localist perspective.

\section{Preliminaries}\label{sec: prelim}\subsection{Partial Information Decomposition}\label{sec: pid}
Let us consider a set of $d$ source variables $S = \{S_1, \ldots, S_d\}$ and a target variable $T$. The Partial Information Decomposition framework provides a principled way to dissect the mutual information between the set $S$ and $T$,  $I(S; T)$, by
splitting it into components that represent distinct modes of information sharing: the unique, redundant, and synergistic contributions from the sources to the target~\citep{williams2010nonnegative}.
At the core of the PID formalism lies a combinatorial object, the redundancy lattice~\citep{williams2010nonnegative}. 
A lattice is a set equipped with a partial order such that every pair of elements in the set has a well-defined meet and join~\citep{rota1964foundations,birkhoff1940lattice}.
In the PID setting, the underlying set $\mathcal{A}(S)$ is the collection of antichains drawn from the powerset of the powerset of $S$. Formally, each element $\alpha \in \mathcal{A}(S)$ is a collection of non-empty subsets $A_i \subseteq S$ such that no $A_i \in \alpha$ is a subset of another~\citep{williams2010nonnegative}:
\begin{equation*}
    \mathcal{A}(S)=\left\{\alpha \in \mathcal{P}_1\left(\mathcal{P}_1(S)\right): \forall A_i, A_j \in \alpha, A_i \not \subset A_j\right\}\, ,
\end{equation*}
where $\mathcal{P}_1(S)$ denotes the powerset of $S$ excluding the empty set.
The partial order $\preceq$ on $\mathcal{A}(S)$ is defined by 
\begin{equation*}
    \alpha \preceq \beta\quad \Longleftrightarrow \quad\forall B \in \beta,\, \exists A \in \alpha \,\text{ such that } A \subseteq B\,.
\end{equation*}
This ordering reflects containment: $\alpha \preceq \beta$ if every subset in $\beta$ contains at least one subset in $\alpha$ so that the information represented by $\beta$ subsumes that represented by $\alpha$.

The key link between the redundancy lattice and PID is that any element $\beta \in \mathcal{A}(S)$ is associated with a redundancy measure $I_\cap(\beta; T)$ that quantifies the amount of information about $T$ that is redundantly available from any ${B}\in \beta$.
Crucially, $I_\cap(\beta; T)$
can be expressed as a sum of partial information atoms, $\Pi_{R}(\alpha; T)$, which represent the irreducible units of shared information and can be viewed as the Zeta function in the M\"obius inversion theorem~\citep{rota1964foundations}:
\begin{equation}
\label{eq:redundancy_measure}
    I_\cap(\beta; T) = \sum_{\alpha \preceq \beta} \Pi_R(\alpha; T)\,.
\end{equation}
Inverting Equation~\eqref{eq:redundancy_measure} via  M\"obius recovers the partial information atoms from the redundancy function~\citep{williams2010nonnegative}.

In the case of two sources, $S = \{S_1, S_2\}$, the redundancy lattice has four elements:  $\{\{S_1\}, \{S_2\}\}$, $\{\{S_1\}\}$, $\{\{S_2\}\}$ and $\{\{S_1, S_2\}\}$,
and the four corresponding partial information atoms can be interpreted as redundant, unique and synergistic components:
\begin{align*}
    I(S_1, S_2; T) = 
    I_\cap (\{\{S_1,S_2\}\},T)  &= \mathrm{Red}(\{\{S_1\}, \{S_2\}\};T) + \mathrm{Uniq}(\{\{S_1\}\};T)  \nonumber \\
    &\quad + \mathrm{Uniq}(\{\{S_2\}\};T) + \mathrm{Syn}(\{\{S_1, S_2\}\};T)\,.
    \label{eq:2_variable_PID}
\end{align*}
$\mathrm{Red}$ denotes the redundant contribution of the two sources; $\mathrm{Uniq}$, information uniquely attributable to one source; and $\mathrm{Syn}$, synergistic information available only when both are taken together. 

\underline{\textit{Remark}:} The self-redundancy axiom~\citep{williams2010nonnegative} implies that the mutual information between sources and target is equal to the redundancy measure of the maximum element in the redundancy lattice, i.e.,
$I(S,T) = I_\cap(\{\{S_1,\ldots,S_d\}\},T)\,$.

\underline{\textit{Remark}:} 
For $d>2$, the situation is more intricate. 
Although inverting Eq.~\eqref{eq:redundancy_measure} via  M\"obius inversion allows in principle the recovery of the partial information atoms from the redundancy measure, the size of the redundancy lattice grows according to the Dedekind numbers~\citep{jakel2023computation}, which makes computation infeasible beyond $d>3$. The structures of the redundancy lattice for $d=2$ and $d=3$ are shown in Figure~\ref{fig:lattice} in Appendix~\ref{app: pid}.

\subsection{Bayesian network}
A natural choice to encode the interactions within a set of variables is through a graph $\mathcal{G} = (V, E)$, where the set of vertices $V$ is associated with the variables and the set of edges $E \subseteq V \times V$ capture the interactions between variables. When $\mathcal{G}$ is directed, an edge is an ordered pair $(X_i, X_j)\in E$, which indicates that $X_i$ is a parent of $X_j$, or equivalently, $X_j$ is a child of $X_i$. The set of parents (children) of node $X_i$ in the graph is denoted as $\text{pa}(X_i)$ ($\text{ch}(X_i)$).
A directed acyclic graph (DAG) is a directed graph  with no sequence of directed edges that forms a closed loop.

Causal structural graphs are formulated as 
Bayesian networks~\citep{bishop2007}, i.e., DAGs where the $d$ vertices represent random variables $\{X_1,\ldots, X_d\}$ with a distribution function
 $\Prob(X_1, \ldots, X_d)$. Then $\Prob$ can be factorized as:
\begin{equation*}
\Prob(X_1, \ldots, X_d) = \prod_{i=1}^d \Prob(X_i \mid \text{pa}(X_i))\,.
\end{equation*}
See Figure~\ref{fig:example}a for an example.
The d-separation criterion then provides a systematic method for translating the Bayesian network into a set of conditional independence properties defined by the DAG structure~\citep{bishop2007}.

While Bayesian networks are suitable for representing pairwise, directed relationships, their intrinsic edge-based topology is not designed to explicitly encode higher-order interactions. To naturally capture these complex, higher-order directed and undirected relationships, we must turn to a more expressive causal graphical representation, the Bayesian hypergraph.

\subsection{Bayesian hypergraph}
A hypergraph $\mathcal{H} = (V, \mathcal{E})$ is an extension of pairwise graphs where each hyperedge $\varepsilon\in\mathcal{E}$ is associated with a subset of the nodes in $V$ with arbitrary cardinality.  In a directed hypergraph, every hyperedge is an ordered pair of subsets $\varepsilon=(T(\varepsilon), H(\varepsilon))$ such that the tail $T(\varepsilon)$ and head $H(\varepsilon)$ do not overlap, and the direction is from tail to head.
(Note that the head or tail could be the empty set.)
Then we say that $t \in T(\varepsilon)$ is a parent of $h\in H(\varepsilon)$, or, equivalently, $h$ is a child of $t$. Furthermore, two vertices $X_i, X_j \in V$ are said to be co-heads (co-tails) when there exists a directed hyperedge $\varepsilon$ whose head (tail) contains both vertices.
A partially directed cycle in $\mathcal{H}$ is a sequence of vertices $(X_1, X_2, \ldots, X_k)$ with $X_{k+1}=X_1$
such that each $X_i$ is either a co-head or a parent 
of its successor in the sequence, $X_{i+1}$, for all $1 \le i \le k$.
A directed acyclic hypergraph is a directed hypergraph with no partially directed cycles.
A hyperedge $\varepsilon_1$ is contained in hyperedge $\varepsilon_2$ if $T(\varepsilon_1)\cup H(\varepsilon_1) \subset T(\varepsilon_2)\cup H(\varepsilon_2)$. 
When a hyperedge is not contained by any other hyperedges it is called irreducible.  See~\citet{javidian2020hypergraph} for definitions.

\begin{figure}[t]
    \centering
    \includegraphics[width=0.55\linewidth]{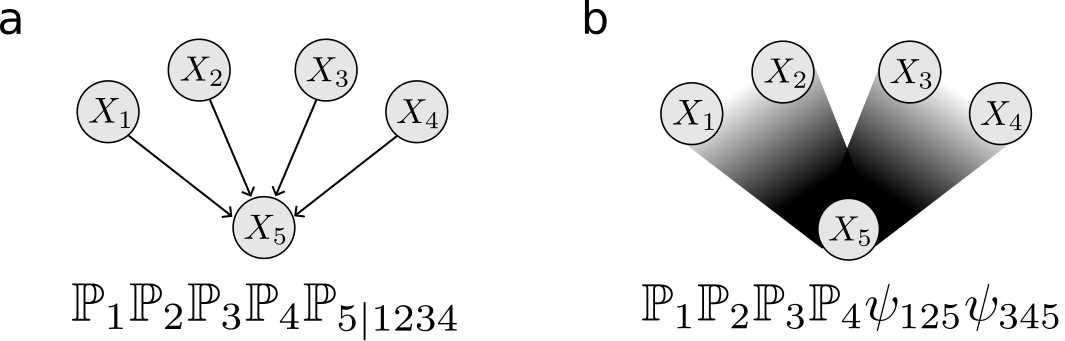}
    \caption{\textbf{Factorization of distributions according to Bayesian network and Bayesian hypergraphs.} (a) A Bayesian network where $X_1$, $X_2$, $X_3$ and $X_4$ are the parents of $X_5$. 
    (b) A Bayesian hypergraph where $X_1$ and $X_2$ (similarly $X_3$ and $X_4$) are co-parents of $X_5$. Conditioning on $X_5$ only induces dependence between $\{X_1,X_2\}$ and $\{X_3, X_4\}$ whereas in (a) conditioning on $X_5$ induces dependence among all parent pairs. The factorizations (up to a normalizing constant) are given below each structure. Hyperedges are shaded directionally from (co-)tail (light) to (co-)head (dark). }
    \label{fig:example}
\end{figure}

A Bayesian hypergraph is a probabilistic model defined via a directed acyclic hypergraph, where the nodes represent random 
variables and the hyperedges encapsulate dependencies between sets of variables~\citep{javidian2020hypergraph}. The Bayesian hypergraph provides a principled integration of directed and undirected relationships through a two-level hierarchical factorization structure, as follows. 
First, the variables are coarse-grained into \textit{chain components} $\tau$; these are equivalence classes where $X_i$ and $X_j$ belong to the same $\tau$ if there exists a sequence of distinct variables $(X_i, \ldots, X_j)$ 
such that each consecutive pair 
in the sequence are co-heads.
A canonical DAG is formed by taking the chain components of $\mathcal{H}$ as nodes and a directed edge from a chain component $\tau_1$ to another chain component $\tau_2$ is present if there exists $u\in \tau_1$, $v\in \tau_2$ and a hyperedge $\varepsilon$ such that $u\in T(\varepsilon)$ and $v\in H(\varepsilon)$.
The joint distribution then factorizes similarly to a Bayesian network over the set of chain components, $\mathcal{D}$: 
\begin{align*}
\Prob(X_1, \ldots, X_d) = \prod_{\tau \in \mathcal{D}} \Prob(X_\tau \mid X_{\text{pa}(\tau)})\,,
\end{align*}
where $X_\tau$ denotes the set of variables in chain component $\tau$, and $\text{pa}(\tau)$ represents the parents of chain component $\tau$ in the canonical DAG of the hypergraph. 
Second, within each $\tau$, the conditional distribution functions can be factorized into non-negative potential functions, akin to those in Markov random fields~\citep{bishop2007}, each specified by an irreducible hyperedge. This leads to the following factorization over the set of irreducible hyperedges in $\tau$, $\mathcal{M}_\tau$: 
\begin{align*}
\Prob(X_\tau \mid X_{\text{pa}(\tau)}) \propto 
\prod_{\varepsilon \in \mathcal{M}_\tau} \psi_\varepsilon(X_\varepsilon)\,,
\end{align*}
where 
$\psi_\varepsilon(X_\varepsilon)$ denotes the potential function associated with the irreducible hyperedge $\varepsilon$.
An example of Bayesian hypergraph with its factorization is shown in Figure~\ref{fig:example}b.

\textit{\underline{Remark}:}
The patterns of conditional independence in Bayesian hypergraphs differ from Bayesian networks.
In Bayesian networks, conditioning on a child induces conditional dependence among all parents (Fig.~\ref{fig:example}a). This is not necessarily the case in Bayesian hypergraphs since conditioning on the child activates dependence only between co-tails within the same hyperedge leaving other parent pairs conditionally independent (Fig.~\ref{fig:example}b). Hence Bayesian hypergraphs are strictly more expressive than Bayesian networks in both their probabilistic factorization patterns and their graphical representations.

\textit{\underline{Remark}:}
Another key advantage of directed hypergraphs is that the co-head relationship provides a natural framework for accommodating indeterminacy in the causal direction between variables. When conditional independence queries reveal dependencies without clear directional information, hypergraphs can represent these relationships through co-head structures that capture the symmetric aspects of the interaction while preserving the overall causal hierarchy.

\section{PID Signatures of Causal Structures}\label{sec: method}
In this section, we first introduce theoretical conditions to establish a bidirectional correspondence between causal roles and information-theoretic signatures for Bayesian networks. 
We then generalize these results to Bayesian hypergraphs, where the richer structural relationships enable more nuanced PID signatures for causal configurations. The proofs can be found in Appendix~\ref{app: proof}.
We start by defining our problem, desiderata\footnote{Desiderata are properties desired for the PID measures to fit in our causal discovery framework, whereas the assumptions are precise mathematical statements we accept as true to prove the theorems.} and assumptions, and establishing our theoretical building blocks.

\begin{desideratum}[Non-negative PID atoms]\label{des: nonnegative red}
It is desirable that the choice of redundancy measure leads to non-negative atoms for all elements of the redundancy lattice:  $\Pi_R(\alpha;T) \geq 0, \forall \alpha \in \mathcal{A}(S)$.
\end{desideratum}
As there is indeterminacy in the definition of the redundancy measure in PID~\citep{lizier2018information, rauh2014reconsidering}, Desideratum~\ref{des: nonnegative red} ensures that each lattice element contributes a meaningful (non-negative) amount of information to the PID decomposition (Eq.~\eqref{eq:redundancy_measure}). 
For bivariate PID (i.e., $d=2$ source variables), it is possible to decompose $I(S_1, S_2; T)$ into \textit{non-negative} unique, synergistic, and redundant components following \citet{e16042161}.
For $d>2$, however, \citet{lyu2025multivariate} demonstrated that no redundancy measure can always guarantee non-negativity of all PID components while satisfying other intuitive desiderata, e.g., independent additivity guaranteeing that two independent variables have zero redundant information.

To circumvent this difficulty, in this paper we employ a  formalism based on bivariate PID but apply it to systems involving $d$ variables by building `supervariables' that contain the complement set of any two variables of interest. This allows us to exploit the well-defined bivariate PID components (unique, synergistic, redundant) of a variable but now with respect to the rest of variables.  
For a set of variables $V$ with target variable $X_k \in V$ and source variable $X_i \in V \setminus \{X_k\}$, we thus define the shorthand notation:
\begin{equation*}
    \Pi_R^{V'}(\{\{X_i\}\}; X_k) \equiv \Pi_R^{\{X_i, X'\}}(\{\{X_i\}\}; X_k)\,, 
\end{equation*}
where $X' = V \setminus \{X_i, X_k\}$ is the supervariable containing all variables except $X_i$ and the target $X_k$, and  $V' = \{X_i, X'\}$ is the bivariate source set for the PID decomposition.
By grouping the rest of variables into $X'$, this notation ensures that the unique information represents only what $X_i$ can provide that is truly unavailable from any combination of the remaining variables. This formulation enables PID causal formalization for $d$ variables while maintaining the theoretical guarantees of bivariate PID.
Note that increasing the source variables from $S$ to $S' \supset S$ increases the lattice elements that precede any particular element in the lattice ordering; hence it can occur that 
previously unique information is reclassified as redundant in the expanded source set.
Therefore, our notation for the PID components includes explicitly the source set $S$,  i.e., $\Pi^{S}_{R}(\alpha; T)$.

First, we establish a fundamental result on the relationship between conditional mutual information and unique information, namely, that the vanishing of  conditional mutual information implies zero unique information. 
\begin{theorem}[Conditional independence and unique information]
\label{thm:conditional_independence_unique_info}
Let $C \subseteq V$ be a conditioning set, and $X_i, X_j \in V  \setminus C$.
Given Desideratum~\ref{des: nonnegative red}, if there exists $C$ such that $I(X_i; X_j \mid C) = 0$, then $X_j$ has zero unique information when $X_i$ is set as the target variable, i.e., $\Pi^{V'}_R(\{\{X_j\}\}; X_i) = 0\,$.
\begin{proof}
    See Section~\ref{proof of theorem 1} of the appendix.
\end{proof}
\end{theorem}
This theorem is a crucial building block for establishing the connection between direct causal neighbors and unique information patterns. For any variable that is not a direct causal neighbor of the target, we can always construct some conditioning set that renders their conditional mutual information zero, and through this theorem we can then establish that such variables contribute zero unique information to the target.

Another key condition that should be satisfied is the monotonicity of unique information.
\begin{desideratum}[Monotonicity of unique information]\label{def:unique_info_monotonicity}
Let $X_1, X_2$ be the source variables, and  for any additional variable $K$, let $X_2' = \{X_2\} \cup \{K\}$ be the supervariable that combines $K$ and $X_2$. 
We then require:
 $   \Pi^{\{X_1,X_2'\}}_R\left(\{\{X_1\}\}; T\right) \leq \Pi^{\{X_1,X_2\}}_R\left(\{\{X_1\}\}; T\right)$\,.
\end{desideratum}
In other words, Desideratum~\ref{def:unique_info_monotonicity} formalizes the intuitive notion that, in bivariate PID, augmenting one source variable with additional information cannot increase the unique information contribution of the other source variable.
The supervariable $X_2'$ contains all the information that $X_2$ originally had plus additional information from $K$. Therefore, any information previously unique to $X_1$ (relative to $X_2$) may now be partially or fully captured by $X_2'$, but no new unique information for $X_1$ should emerge from this augmentation. This monotonicity property ensures that the notion of uniqueness remains coherent as we expand the information available to competing sources.
Note how the use of supervariable $X'_2$ allows us to remain in a bivariate PID framework.

To establish the bidirectional correspondence between causal roles and their PID signatures, we also require the following constraint on the joint probability distribution over all variables.
\begin{assumption}[Persistent relevance]
\label{as:persistent_relevance}
If $I(X_i; T\mid C) > 0$ 
for all conditioning sets $C\subseteq S \setminus \{X_i\}$, 
then $X_i$ must possess positive unique information about target $T$, i.e., $\Pi_R(\{\{X_i\}\}; T) > 0$. 
\end{assumption}
This property addresses pathological cases in PID where a variable might always maintain some dependency with the target, yet contribute zero unique information. 
For example, consider two independent coin flips $X_1$ and $X_2$, and let $X_3=X_1\oplus X_2$ such that $X_1$ and $X_2$ will only have synergistic information about $X_3$ due to the XOR gate. Given another independent coin flip $X_4$, construct new variables $X_1'=\{X_1,X_4\},X_2'=\{X_2,X_4\},X_3'=\{X_3,X_4\}$. Setting $X_3'$ as the target, $X_1'$ and $X_2'$ will only have redundant information and synergistic information, as $X_4$ is perfectly replicated across three variables, which means any information about $X_4$ provided by $X_1'$ can also be provided by $X_2'$, where as information about $X_3$ can only be attained synergistically. Since $X_3$ and $X_4$ together determine $X_3'$,  there is no room for unique information from either variable.

\subsection{PID signatures of causality in Bayesian networks}\label{sec: pid_bn}
 \subsubsection{Direct causal neighbors and unique information}

We now present one of our central results establishing the correspondence between unique information and causal neighbors. 
In this section, we assume the standard faithfulness condition~\citep{spirtes2000causation}, i.e., all statistical independencies in the joint distribution correspond precisely to the d-separations in the underlying Bayesian network.

\begin{theorem}[Causal relationships and unique information]
\label{thm:causal_unique_information}
Under Assumption~\ref{as:persistent_relevance}, for any $X_i$ and $X_j$ in a Bayesian network, 
the unique information of $X_j$ about $X_i$ is positive iff $X_j$ is a parent or child of $X_i$.
\begin{proof}
    See Section~\ref{proof of theorem 2} of the appendix.
\end{proof}
\end{theorem}
This result formalizes our intuition that a parent exerts an influence on the child variable that cannot be derived from any other source, while the child is influenced by the parent variable in ways that reflect information unavailable elsewhere in the system. 
Importantly, Theorem~\ref{thm:causal_unique_information} supports a localist interpretation of causal roles. In contrast to standard causal learning approaches, which proceed by narrowing down causal models through systematic conditional independence queries across the entire variable set, this allows us to focus locally on a  particular variable of interest and identify those variables that contribute unique information to it, which are either its parents or children. 

\underline{\textit{Remark}:} The estimation of redundant information in PID, which depends on multiple variables simultaneously, can be reused across different targets. This shared structure allows the unique information for several variables to be inferred from one redundancy computation, thus reducing computational cost. Such reuse has no analogue in standard causal methods based on conditional independence tests, and suggests new possibilities for more efficient structural discovery algorithms. An example of redundancy reuse is shown in Appendix~\ref{app: red_reuse}.

\subsubsection{Co-parent and synergy}

After identifying the PID signature of parents and children, we now seek to distinguish between these two roles. 
We do this by identifying the characteristic PID signature of causal co-parents and introducing appropriate theoretical conditions to ensure that this identification is bidirectional, i.e., co-parent relationships both imply and are implied by specific PID patterns.

Before proving our result, we state the necessary assumptions. First, we introduce the collider amplification assumption which formalizes the intuition that the collider effect should always increase mutual information between co-parents, regardless of whether they have nonzero unconditional mutual information beforehand. 
\begin{assumption}
[Collider amplification]
\label{as:collider_amplification}
For every collider $X_i \to X_j \leftarrow X_k$ in a Bayesian network, we assume that the following holds:
$I(X_i; X_k \mid X_j) > I(X_i; X_k)\,.$
\end{assumption}

\underline{\textit{Remark}:} Clearly, the pure V-structure satisfies $I(X_i; X_k) = 0$ by d-separation and Assumption~\ref{as:collider_amplification} holds directly. Yet additional dependencies between $X_i$ and $X_k$ induced by other variables in the graph might result in $I(X_i; X_k) > 0$. When this happens, d-separation alone cannot specify by itself the relative magnitude of $I(X_i; X_k)$ versus $I(X_i; X_k \mid X_j)$.

Second, a key property that should be satisfied by any reasonable redundancy measure is that adding information to some variable will not reduce the value of any PID atoms that contains it:
\begin{desideratum}
[Information monotonicity]\label{des:info_monotonicity}
Given $X \in S$ and another disjoint set of random variables $K \cap S = \emptyset$, let: (i) $X'$ be the supervariable that combines $X$ and $K$; (ii) $S'$ be the source set obtained by replacing $X$ with $X'$; (iii) $\alpha'$ be the lattice element in the bivariate redundancy lattice obtained from $\alpha$ by replacing $X$ with $X'$.
We require
$
    \Pi^{S'}_R(\alpha') \geq \Pi^S_R(\alpha)
$
for all $\alpha$ that contains a subset that contains $X$.
\end{desideratum}
Since $X'$ contains all the information that $X$ has (and potentially more from $K$), replacing $X$ with $X'$ in any lattice element should preserve or increase the corresponding PID component.
This property ensures that when we replace $X$ with the augmented variable $X'$, any synergistic, redundant, or unique information will maintain access to the original information in $X$ while potentially gaining additional information from $K$. 

We now present our result that identifies co-parents using synergy. This result complements Theorem~\ref{thm:causal_unique_information}, which linked positive unique information to parents and children. 
\begin{theorem}[Co-parent identification via synergy]\label{thm:coparent_detection}
Under Assumptions~\ref{as:persistent_relevance} and \ref{as:collider_amplification}, 
for $X_i, X_k \in V$, $X_k$ is a co-parent of $X_i$ if and only if:
\begin{enumerate}[label=(\roman*)]
    \item $X_k$ has zero unique information about $X_i$: $\Pi^{V'}_R(\{\{X_k\}\}; X_i) = 0$;
    \item There exists $X_j \in V$ with $\Pi_R^{V'}(\{\{X_j\};X_i\})>0$ such that $X_k$ and $X_j$ have positive synergy about $X_i$: $\Pi^{\{X_j,X_k\}}_R(\{\{X_j, X_k\}\}; X_i) > 0$, where $X_j$ is the joint child of $X_k$ and $X_i$.
\end{enumerate}
Note that the zero unique information is relative to the set of all variables $V$, while positive synergy is only relative to $\{X_j,X_k\}$.
\begin{proof}
    See Section~\ref{proof of theorem 3} of the appendix.
\end{proof}
\end{theorem}
Identifying the co-parent PID signature enables us to confirm whether a variable $X_j$ is a child of the target variable $X_i$. Conversely, to verify that $X_j$ is a parent of $X_i$, we reverse the roles by setting $X_j$ as the target and attempting to identify a co-parent of $X_j$ with $X_i$ serving as their joint child. 

\underline{\textit{Remark}:} While the identification of co-parents requires finding pairs of variables corresponding to the collider and the co-parent, we can narrow down the set of possible colliders through Theorem~\ref{thm:causal_unique_information}, which drastically improves efficiency if the graph is sparse. In particular, we only have to check for co-parent structures between two variables $X_i$ and $X_j$ when their direct neighborhoods $N(X_i)$ and $N(X_j)$ intersect since the collider has unique information to both $X_i$ and $X_j$.

By combining Theorems~\ref{thm:causal_unique_information}~and~\ref{thm:coparent_detection}, which allow direct causal neighbor and co-parent identification, we provide a systematic causal discovery procedure for Bayesian networks based on PID.
\begin{proc}[PID-based causal discovery for Bayesian networks]\label{thm:pid_bayesnet_discovery}
Under Assumptions~\ref{as:persistent_relevance} and \ref{as:collider_amplification} and Desiderata~\ref{des: nonnegative red}--
\ref{des:info_monotonicity}
the causal structure of a Bayesian network can be identified through the following three-step procedure:
\begin{enumerate}
    \item \textbf{Direct Causal Neighbor Identification.} For each $X_i \in V$, set $X_i$ as the target for PID analysis and identify all variables with positive unique information about $X_i$ 
    (Theorem~\ref{thm:causal_unique_information}).
    \item \textbf{Directional Confirmation via Co-parent Detection.} For each $X_i \in V$ and each of its direct causal neighbors $X_j \in N(X_i)$, attempt to identify a co-parent variable such that unique information about $X_i$ is zero and synergistic information with $X_j$ is positive (Theorem \ref{thm:coparent_detection}).
    The identification of such a co-parent $X_k$ confirms that $X_j \in \text{ch}(X_i)$ (equivalently, $X_i \in \text{pa}(X_j)$), thereby establishing the causal direction $X_i \to X_j$.

    \item \textbf{Remaining Direction Resolution.} After co-parent relationships have been used to direct edges, apply remaining heuristics (e.g., acyclicity constraints) to complete any indeterminate causal relationships that could not be resolved through co-parent identification.
\end{enumerate}
\end{proc}

\subsection{PID signatures in Bayesian hypergraphs}\label{sec: pid_BH}
In this section, we extend our results linking causality and PID to Bayesian hypergraphs. Crucially, we identify the PID signatures of variables within individual hyperedges that  capture both the directed relationships between tails and heads, as well as the symmetric relationships among co-heads and co-tails. Here we only consider interactions that can be faithfully represented by a Bayesian hypergraph, such that all statistical independencies correspond exactly to the conditional independence properties dictated by the hypergraph factorization.

\underline{\textit{Remark}:}
As Bayesian hypergraphs generalizes Bayesian networks, being faithful to a Bayesian network implies faithfulness to a Bayesian hypergraph, but the reverse does not hold. Under standard faithfulness assumptions, neither Bayesian networks nor Bayesian hypergraphs are fully designed to represent strict logical XORs~\citep{marx2021weaker}. 

\underline{\textit{Remark}:}
To illustrate interactions that standard Bayesian networks fail to faithfully represent but Bayesian hypergraphs can, consider the following example of a partial collider. Suppose $X_1$, $X_2$, $X_3$, and $X_4$ are independent coin flips, and $X_5 = (Y_1, Y_2)$ where $Y_1 = X_1 \text{ OR } X_2$ and $Y_2 = X_3 \text{ OR } X_4$. Then, conditioning on $X_5$ activates both colliders, rendering $X_1$ dependent on $X_2$, and $X_3$ dependent on $X_4$. 
Crucially, the pair $\{X_1, X_2\}$ remains conditionally independent of $\{X_3, X_4\}$ given $X_5$. A standard Bayesian network cannot faithfully capture this structure; conditioning on the common child $X_5$ would incorrectly induce dependence among all parent pairs, resulting in the structure shown in Fig.~\ref{fig:example}a. A Bayesian hypergraph, however, correctly encodes these as two disjoint causal mechanisms meeting at $X_5$, preserving the conditional independence between the distinct groups through its structured factorization, as illustrated in Fig.~\ref{fig:example}b.

Similar to Bayesian networks, we also require the assumption of collider amplification for hypergraphs, with the notion of co-parents replaced by co-tails.
\begin{assumption}[Hypergraph collider amplification]\label{as:hypergraph_collider_amplification}
Given a directed hyperedge $\varepsilon$ in Bayesian hypergraph $\mathcal{H}$ such that $X_i, X_k \in T(\varepsilon)$ and $X_j \in H(\varepsilon)$, the following holds:
\begin{equation*}
    I(X_i; X_k \mid X_j) > I(X_i; X_k)\,.
\end{equation*}
\end{assumption}
The hypergraph collider amplification property states that the collider effect only activates for variables that are co-tails within the same directed hyperedge. This restriction is necessary due to the factorization structure of directed hypergraphs.
For instance, $X_1$ and $X_3$ are not co-tails of any hyperedge in the Bayesian hypergraph in Figure~\ref{fig:example}b; hence conditioning on $X_5$ does not create conditional dependence between $X_1$ and $X_3$, as they belong to separate potential functions $\psi_{125}$ and $\psi_{345}$. This is in contrast to the Bayesian network in Figure~\ref{fig:example}a, where conditioning on $X_5$ creates conditional dependence between any pair of parent variables. Therefore Assumption~\ref{as:collider_amplification} requires variables to be joint causes (co-tails) of the same effect (head) within a single directed hyperedge.

Next, we introduce our theorem on the relationship between unique information and causal neighbors in hypergraphs. This is the Bayesian hypergraph parallel of Theorem \ref{thm:causal_unique_information}.

\begin{theorem}[Hypergraph unique information characterization]\label{thm:hypergraph_unique_info}
Under Assumption~\ref{as:persistent_relevance}, for any variable $X_i$ and $X_j$ in a Bayesian hypergraph, the unique information of $X_j$ about $X_i$ is positive iff $X_j$ is a parent, child, or co-head of $X_i$.
\begin{proof}
    See Section~\ref{proof of theorem 4} of the appendix.
\end{proof}
\end{theorem}
In Bayesian hypergraphs, the variables with positive unique information correspond not only to parents and children, but also to co-heads because the factorization of Bayesian hypergraphs contains both asymmetric causal relationships (parent/child) and symmetric relationships (co-heads).

We now present the Bayesian hypergraph analogue of Theorem~\ref{thm:coparent_detection}. For hypergraphs, the collider effect only activates among co-tails, hence we replace co-parents in Theorem~\ref{thm:coparent_detection} with co-tails.
\begin{theorem}[Hypergraph co-tail identification via synergy]\label{thm:hypergraph_cotail_detection}
Under Assumptions~\ref{as:persistent_relevance} and \ref{as:hypergraph_collider_amplification}, $X_k$ is a co-tail of variable $X_i$  
if and only if:
\begin{enumerate}[label=(\roman*)]
\item $X_k$ has zero 
unique information about $X_i$: $\Pi^{V'}_R(\{\{X_k\}\}; X_i) = 0$;
\item There exists $X_j \in V$ with $\Pi^{V'}_R(\{\{X_j\}\}; X_i) > 0$ such that $X_k$ and $X_j$ have positive synergy about $X_i$: $\Pi^{\{X_j,X_k\}}_R(\{\{X_j, X_k\}\}; X_i) > 0$.
\end{enumerate}
\begin{proof}
    See Section~\ref{proof of theorem 5} of the appendix.
\end{proof}
\end{theorem}

By combining our results on Bayesian hypergraph unique information and co-tail identification (Theorems~\ref{thm:hypergraph_unique_info}~and~\ref{thm:hypergraph_cotail_detection}), we can pin down the PID signature of a hyperedge as follows:
\begin{theorem}[PID signatures of hyperedges]\label{thm:hyperedge_pid_signature}
Under Assumptions~\ref{as:persistent_relevance}~and~\ref{as:hypergraph_collider_amplification}, for any directed hyperedge $\varepsilon$ with head $H(\varepsilon)$ and tail $T(\varepsilon)$, the following PID signatures hold:
\begin{enumerate}[label=(\roman*)]
\item For every $X_i \in T(\varepsilon)$ and $X_j \in H(\varepsilon)$, any co-tail $X_k \in T(\varepsilon) \setminus \{X_i\}$ has positive synergy with $X_j$ about $X_i$:
$
    \Pi^{\{X_j,X_k\}}_R(\{\{X_j, X_k\}\}; X_i) > 0\,.
$
\item For every $X_i \in T(\varepsilon)$ and $X_j \in H(\varepsilon)$, the head $X_j$ has positive unique information about $X_i$:
$
    \Pi^{V'}_R(\{\{X_j\}\}; X_i) > 0\,.
$
\item Any pair of co-heads $X_j, X_k \in H(\varepsilon)$
have positive unique information about each other:
$
    \Pi^{V'}_R(\{\{X_j\}\}; X_k) > 0\,\, \text{ and }\,\, \Pi^{V'}_R(\{\{X_k\}\}; X_j) > 0\,.
$
\end{enumerate}
\begin{proof}
    See Section~\ref{proof of theorem 6} of the appendix.
\end{proof}
\end{theorem}

While each hyperedge implies particular PID signatures, a particular PID signature may be compatible with multiple combinations of hyperedges. To resolve this ambiguity, we introduce the notion of maximal hyperedge.
\begin{definition}[Maximal hyperedge]\label{def:maximal_hyperedge}
A directed hyperedge $\varepsilon^* \in \mathcal{E}$ is called maximal if extending either its head $H(\varepsilon^*)$ or tail $T(\varepsilon^*)$ would cause it to violate the PID signature properties in Theorem~\ref{thm:hyperedge_pid_signature}. 
Formally, $\varepsilon^*$ is maximal if there does not exist a hyperedge $\varepsilon'$ with either:
$T(\varepsilon^*) \subset T(\varepsilon')$ and $H(\varepsilon^*) = H(\varepsilon')$, or
$T(\varepsilon^*) = T(\varepsilon')$ and $H(\varepsilon^*) \subset H(\varepsilon')$
such that $\varepsilon'$ also satisfies the PID signature properties in Theorem~\ref{thm:hyperedge_pid_signature}.
\end{definition}
\begin{figure}[t]
    \centering
    \includegraphics[width=0.33\linewidth]{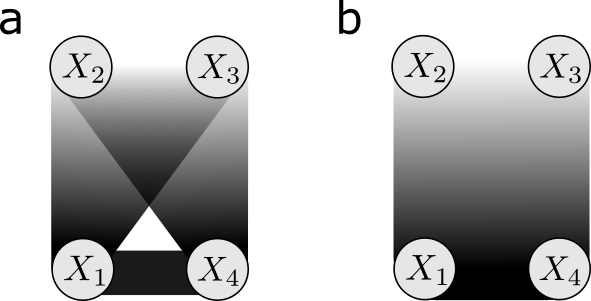}
    \caption{\textbf{Example of maximal hyperedge construction.} The Bayesian hypergraph with three hyperedges in (a) has the same conditional independence properties as the one in (b). Yet (b) provides a more parsimonious representation using the maximal hyperedge construction.
 }
    \label{fig:maximal hyperedge}
\end{figure}
The maximal hyperedge construction provides a canonical representation that consolidates redundant hyperedge specifications. For example, consider the Bayesian hypergraph in Figure~\ref{fig:maximal hyperedge} with three hyperedges: $\varepsilon_1$ with $T(\varepsilon_1) = \{X_1, X_2\}$ and $H(\varepsilon_1) = \{X_3\}$; $\varepsilon_2$ with $T(\varepsilon_2) = \{X_1, X_2\}$ and $H(\varepsilon_2) = \{X_4\}$; $\varepsilon_3$ with $T(\varepsilon_3) = \emptyset$ and $H(\varepsilon_3) = \{X_3, X_4\}$.
These can be absorbed into a single maximal hyperedge: $\varepsilon^*$ with $T(\varepsilon^*) = \{X_1, X_2\}$ and $H(\varepsilon^*) = \{X_3, X_4\}$, which results in the same conditional independence properties while providing a more parsimonious representation of the hypergraph structure.

Based on Theorem~\ref{thm:hyperedge_pid_signature} and Definition~\ref{def:maximal_hyperedge}, we develop a  systematic procedure for Bayesian hypergraph causal discovery that leverages information-theoretic patterns to identify canonical causal representations.
\begin{proc}[PID-based causal discovery for Bayesian hypergraph]\label{thm:pid_hypergraph_discovery}
Under Assumptions~\ref{as:persistent_relevance} and 
\ref{as:collider_amplification} and Desiderata~\ref{des: nonnegative red}-- 
\ref{des:info_monotonicity},
the following three-step procedure identifies the Bayesian hypergraph structure:
\begin{enumerate}
\item \textbf{Candidate Search}: Search for subsets $T \subseteq V$ (potential tails) and $H \subseteq V$ (potential heads) with $T \cap H = \emptyset$ that satisfy the hyperedge PID signature properties of Theorem \ref{thm:hyperedge_pid_signature}.

\item \textbf{Maximal Extension}: For each candidate hyperedge $(T, H)$ that satisfies the PID signature, expand both $T$ and $H$ by adding variables while maintaining the hyperedge signature properties, continuing until no further expansion is possible without violating the signature.

\item \textbf{Canonical Selection and Iterative Construction}: Select the maximal hyperedge $(T^*, H^*)$ 
according to Definition \ref{def:maximal_hyperedge}. Continue this search process iteratively until no additional maximal hyperedges can be identified. The collection of maximal hyperedges specifies the causal structure of the underlying Bayesian hypergraph.
\end{enumerate}
\end{proc}

\section{Related Works}\label{sec: related_works}

An important branch of causal discovery stems from algorithmic information theory (AIT)~\citep{mian2021discovering,mameche2022discovering,janzing2010causal,kaltenpoth2019we}, which utilizes the Kolmogorov complexity, an inherently noncomputable metric, to identify edge directions via the algorithmic Markov conditions. In contrast, our framework is grounded in Shannon information theory, leveraging the fine-grained components of mutual information to identify the causal roles of nodes, a formulation that can be practically evaluated using existing estimators~\citep{venkatesh2023gaussian,pakman2021estimating,barret2015exploration}. While a primary focus of AIT-based methods is to pinpoint the true causal graph from among the DAGs within a Markov equivalence class, our PID-based method does not attempt to break Markov equivalence. Furthermore, PID's higher-order nature enables causal discovery in more expressive Bayesian hypergraphs, extending beyond the pairwise limitations of current AIT methods.

\section{Discussion}\label{sec: discussion}

Our theoretical results establish a direct correspondence between PID signatures and structural causal configurations. This provides a localist perspective that links higher-order information-theoretic measure to structural and causal roles, offering insights that extend beyond standard causal discovery procedures. In particular, we show that these correspondences extend naturally to Bayesian hypergraphs, yielding an information-theoretic characterization of higher-order causal structure.

Several challenges and limitations remain. Because PID ultimately encodes conditional mutual information, it cannot distinguish causal structures that are observationally Markov-equivalent. A natural direction for future work is to develop an AIT-formulated redundancy measure satisfying our desiderata so that our framework can be readily extended to uncover the true causal graph. Furthermore, although our bivariate reduction avoids the combinatorial growth of full multivariate PID, the practical use of our framework still depends on efficient estimation of unique and synergistic information. Existing PID measures often involve high-dimensional optimization or rely on restrictive parametric assumptions, limiting their scalability and applicability. Developing estimation strategies that enable reliable PID-based inference from finite samples is therefore an important direction for future work. Another natural extension is to broaden our structural correspondence beyond Bayesian networks and hypergraphs to more expressive causal frameworks, thereby expanding the scope of the localist perspective developed here.

\acks{RP was supported by the Deutsche Forschungsgemeinschaft (DFG, Project-ID 424778381, TRR 295).}

\bibliography{references}

\newpage
\appendix

\section{Additional Information}
\subsection{Redundancy Lattices}\label{app: pid}.
\begin{figure}[h]
    \centering
    \includegraphics[width=0.55\linewidth]{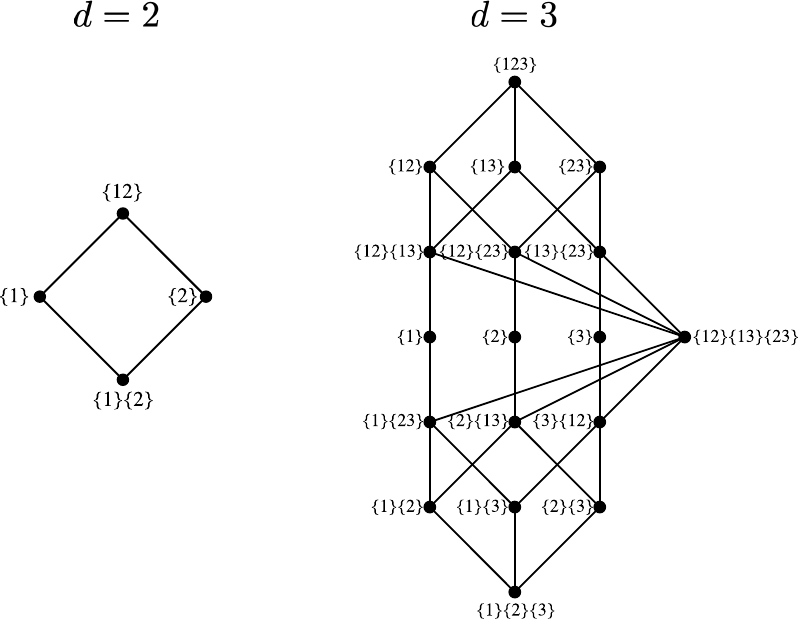}
    \caption{\textbf{Hasse diagrams of the redundancy lattice for $d=2$ (left) and $d=3$ (right) source variables.} Each node corresponds to an element of the redundancy lattice, and edges indicate the partial order defined by information containment. For $d=2$, the lattice contains 4 elements, yielding the familiar bivariate PID structure. For $d=3$, the lattice contains 18 elements, illustrating the rapid growth in lattice size as the number of sources increases. In general, the number of lattice elements grows according to the Dedekind numbers, leading to a combinatorial explosion for $d>3$, which motivates the use of bivariate reductions via supervariables in this work.
 }
    \label{fig:lattice}
\end{figure}

\subsection{Redundancy reuse}\label{app: red_reuse}

Suppose that we have three variables $X_1,X_2,X_3$. To identify the causal skeleton of their relationship we try to estimate the unique information, we can do this by utilizing the relationships 
\begin{align*}
    I(X_1;X_3)=\mathrm{Uniq}(X_1;X_3)+\mathrm{Red}(X_1,X_2;X_3)\\
    I(X_2;X_3)=\mathrm{Uniq}(X_2;X_3)+\mathrm{Red}(X_1,X_2;X_3)
\end{align*}
which gives 
\begin{align*}
    \mathrm{Uniq}(X_1;X_3)= I(X_1;X_3)-\mathrm{Red}(X_1,X_2;X_3)\\
    \mathrm{Uniq}(X_2;X_3)= I(X_2;X_3)-\mathrm{Red}(X_1,X_2;X_3)
\end{align*}
In other words, we only have to calculate the redundant information once and we can use that to estimate multiple unique information terms at once by subtracting it from mutual information.

\section{Proofs}\label{app: proof}
\subsection{Proof of Theorem~\ref{thm:conditional_independence_unique_info}}
\label{proof of theorem 1}
\begin{proof}
By using chain rule of conditional mutual information and decomposing each mutual information term into its PID atoms, we have:
\begin{equation}
I(X_j; X_i \mid C) = \sum_{\alpha \in \mathcal{A}} \Pi_{\mathbf{R}}(\alpha; X_i)
\end{equation}
where $\mathcal{A} = \{\alpha \in \mathcal{L} : \alpha \preceq \{\{X_j, C\}\} \text{ and } \alpha \not\preceq \{\{C\}\}\}$.

By assumption, $I(X_i; X_j \mid C)= 0$, so:
\begin{equation}
\sum_{\alpha \in \mathcal{A}} \Pi_{\mathbf{R}}(\alpha; X_i) = 0
\end{equation}

The set $\mathcal{A}$ contains lattice elements that involve $X_j$ but are not entirely determined by $C$. Crucially, this set includes $\{\{X_j\}\}$ since:
\begin{itemize}
\item $\{\{X_j\}\} \preceq \{\{X_j, C\}\}$ (as $\{X_j\} \subseteq \{X_j, C\}$)
\item $\{\{X_j\}\} \not\preceq \{\{C\}\}$ (since $X_j \notin C$, we have $\{X_j\} \not\subseteq C$)
\end{itemize}

Therefore, $\{\{X_j\}\} \in \mathcal{A}$, and the sum includes $\Pi_{\mathbf{R}}(\{\{X_j\}\}; X_i)$.

Since all PID components are non-negative by our constraint on redundancy measures:
\begin{equation}
\Pi_{\mathbf{R}}(\alpha; X_i) \geq 0 \quad \forall \alpha \in \mathcal{L}
\end{equation}

and the sum of non-negative terms equals zero, each individual term must be zero:
\begin{equation}
\Pi_{\mathbf{R}}(\alpha; X_i) = 0 \quad \forall \alpha \in \mathcal{A}
\end{equation}
Therefore, $X_j$ has zero unique information about target $X_i$.
\end{proof}
\subsection{Proof of Theorem~\ref{thm:causal_unique_information}}\label{proof of theorem 2}
\begin{proof}
\textbf{Direction 1} ($\Rightarrow$): If $X_j$ is not a parent or child of $X_i$, then $\Pi_R^{V'}(\{\{X_j\}\}; X_i) = 0$.

We consider two cases:

\textbf{Case 1}: $X_j \notin MB(X_i)$ (not in the Markov blanket of $X_i$).

By the Markov property, variables outside the Markov blanket are conditionally independent of $X_i$ given the Markov blanket:
\begin{equation}
    X_j \indep X_i \mid MB(X_i)
\end{equation}
This gives $I(X_j; X_i \mid MB(X_i)) = 0$. In the bivariate PID decomposition with source variables $\{X_j, MB(X_i)\}$ (treating the Markov blanket as a single variable), this implies:
\begin{equation}
    \Pi_R^{\{X_j, MB(X_i)\}}(\{\{X_j\}\}; X_i) = 0 
\end{equation}

Since $MB(X_i) \subseteq V' = V \setminus \{X_i, X_j\}$, by the unique information monotonicity property (Definition \ref{def:unique_info_monotonicity}):
\begin{equation}
    Pi_R^{V'}(\{\{X_j\}\}; X_i) \leq \Pi_R^{\{X_j, MB(X_i)\}}(\{\{X_j\}\}; X_i) = 0
\end{equation}
Therefore, $\Pi_R^{V'}(\{\{X_j\}\}; X_i) = 0$.

\textbf{Case 2}: $X_j \in MB(X_i)$ but $X_j \notin \text{pa}(X_i) \cup \text{ch}(X_i)$.

Then $X_j$ must be a co-parent of some child of $X_i$. Let $X_k \in \text{ch}(X_i)$ be a common child of $X_i$ and $X_j$, so we have the V-structure $X_i \to X_k \leftarrow X_j$.

We can construct a conditioning set $C$ as follows:
\begin{itemize}
\item Do not include $X_k$ in $C$ (to avoid activating the collider)
\item For any other path between $X_i$ and $X_j$, include appropriate variables in $C$ to block those paths
\end{itemize}

By d-separation, with this choice of $C$:
\begin{equation}
    X_j \indep X_i \mid C
\end{equation}
Since $C \subseteq V \setminus \{X_i, X_j\} = V'$, by the unique information monotonicity property:
\begin{equation}
    \Pi_R^{V'}(\{\{X_j\}\}; X_i) \leq \Pi_R^{\{X_j, C\}}(\{\{X_j\}\}; X_i) = 0 
\end{equation}

Therefore, $\Pi_R^{V'}(\{\{X_j\}\}; X_i) = 0$.

\textbf{Direction 2} ($\Leftarrow$): If $X_j$ is a parent or child of $X_i$, then $\Pi_{\mathbf{R}}(\{\{X_j\}\}; X_i) > 0$.

If $X_j$ is a parent or child of $X_i$, then there is a direct causal edge between them. By faithfulness, this direct causal connection implies $I(X_j; X_i \mid C) > 0$ for all possible conditioning sets $C \subseteq S \setminus \{X_j\}$.  By the persistent relevance property, this implies $\Pi_{\mathbf{R}}(\{\{X_j\}\}; X_i) > 0$.
\end{proof}
\subsection{Proof of Theorem~\ref{thm:coparent_detection}}\label{proof of theorem 3}

\begin{proof}
\textbf{Direction 1} ($\Leftarrow$): If $X_k$ is a co-parent of $X_i$, then the conditions hold.

Zero unique information: Since $X_k$ is not a direct parent or child of $X_i$ (but rather connected through the collider $X_j$), by Theorem \ref{thm:causal_unique_information}, we have $\Pi^{V'}_R(\{\{X_k\}\}; X_i) = 0$.

Positive synergistic information: Consider the V-structure $X_k \to X_j \leftarrow X_i$, since $X_j$ is a child of $X_i$, we have $\Pi_R(\{\{X_j\}\})>0$. By the collider amplification property:
\begin{equation}
    I(X_k; X_i \mid X_j) > I(X_k; X_i)
\end{equation}
From the two-source PID decomposition:
\begin{align}
    I(X_k; X_i \mid X_j) &= \Pi^{\{X_j,X_k\}}_R(\{\{X_k\}\}; X_i) + \Pi^{\{X_j,X_k\}}_R(\{\{X_j, X_k\}\}; X_i)\\
    I(X_k; X_i) &= \Pi^{\{X_j,X_k\}}_R(\{\{X_k\}, \{X_j\}\}; X_i) + \Pi^{\{X_j,X_k\}}_R(\{\{X_k\}\}; X_i)
\end{align}
Substituting into the collider amplification inequality:
\begin{equation}
    \Pi^{\{X_j,X_k\}}_R(\{\{X_j, X_k\}\}; X_i) - \Pi^{\{X_j,X_k\}}_R(\{\{X_k\}, \{X_j\}\}; X_i) > 0
\end{equation}
Since redundant information is non-negative, $\Pi^{\{X_j,X_k\}}_R(\{\{X_k\}, \{X_j\}\}; X_i) \geq 0$, we conclude:
\begin{equation}
    \Pi^{\{X_j,X_k\}}_R(\{\{X_j, X_k\}\}; X_i) > 0 
\end{equation}

\textbf{Direction 2} ($\Rightarrow$): If the conditions hold, then $X_k$ is a co-parent of $X_i$.

We prove this by contradiction. Assume that variable $X_k$ has zero unique information about $X_i$ but is not a co-parent of $X_i$. We will show that there cannot exist any variable $X_j$ with positive unique information such that $X_k$ and $X_j$ have positive synergistic information about $X_i$.

Since $X_k$ has zero unique information about $X_i$, by Theorem \ref{thm:causal_unique_information}, $X_k$ is not a parent or child of $X_i$. Combined with our assumption that $X_k$ is not a co-parent, this implies that $X_k$ is outside the Markov blanket of $X_i$: $X_k \notin MB(X_i)$.

On the other hand, any variable $X_j$ with positive unique information about $X_i$ must be a parent or child of $X_i$ by Theorem \ref{thm:causal_unique_information}, which means $X_j \in MB(X_i)$.

By the Markov property, since $X_k \notin MB(X_i)$:
\begin{equation}
    I(X_k; X_i \mid MB(X_i)) = 0
\end{equation}
This implies that in the bivariate PID decomposition with source variables $\{X_k, MB(X_i)\}$ (treating the Markov blanket as a single variable):
\begin{equation}\label{eq: 17}
    \Pi^{\{X_k, MB(X_i)\}}_R(\{\{X_k, MB(X_i)\}\}; X_i) = 0
\end{equation}
Since $X_j \in MB(X_i)$, by the information monotonicity property (Desideratum~\ref{des:info_monotonicity}), we have:
\begin{equation}\label{eq: 18}
    \Pi^{\{X_k, MB(X_i)\}}_R(\{\{X_k, MB(X_i)\}\}; X_i) \geq \Pi^{\{X_k, X_j\}}_R(\{\{X_k, X_j\}\}; X_i)
\end{equation}
Combining Eq.~\eqref{eq: 17} and \eqref{eq: 18} above we have:
\begin{equation}
    \Pi^{\{X_k, X_j\}}_R(\{\{X_k, X_j\}\}; X_i) = 0
\end{equation}
Therefore, $X_k$ and $X_j$ have zero bivariate synergistic information, contradicting the assumption that there exists positive synergistic information. Hence, $X_k$ must be a co-parent of $X_i$.
\end{proof}
\subsection{Proof of Theorem~\ref{thm:hypergraph_unique_info}}\label{proof of theorem 4}

\begin{proof}
\textbf{Direction 1} ($\Leftarrow$): If $X_j$ is a parent, child, or co-head of $X_i$, then $\Pi^{V'}_R(\{\{X_j\}\}; X_i) > 0$.

\textit{Parents and children:} By the same argument as in \ref{thm:causal_unique_information}, direct causal connections ensure $I(X_j; X_i \mid C) > 0$ for all conditioning sets $C \subseteq V \setminus \{X_j\}$. By persistent relevance, this implies positive unique information.

\textit{Co-heads:} If $X_j \sim_H X_i$, they appear together in the head $H(h)$ of some directed hyperedge $h$. By the hypergraph factorization structure, this creates a direct statistical dependency through the shared potential function $\psi_h$ that cannot be eliminated by conditioning on other variables, ensuring $I(X_j; X_i \mid C) > 0$ for all $C \subseteq V \setminus \{X_j\}$.

\textbf{Direction 2} ($\Rightarrow$): If $X_j$ is not a parent, child, or co-head of $X_i$, then $\Pi^{V'}_R(\{\{X_j\}\}; X_i) = 0$.

\textit{Outside Markov blanket:} If $X_j \notin MB(X_i)$, then by the Markov property: $X_j \indep X_i \mid MB(X_i)$, giving $I(X_j; X_i \mid MB(X_i)) = 0$. By the argument in Theorem \ref{thm:causal_unique_information}, this implies zero unique information.

\textit{Co-tails:} If $X_j$ is a co-tail of $X_i$ (i.e., $X_j \sim_T X_i$ but not co-head), then there exists a conditioning set $C$ that excludes their common head variables but includes appropriate blocking variables for other paths. By the hypergraph Markov properties, this renders $X_j \indep X_i \mid C$, yielding $I(X_j; X_i \mid C) = 0$ and thus zero unique information.
\end{proof}
\subsection{Proof of Theorem~\ref{thm:hypergraph_cotail_detection}}\label{proof of theorem 5}

\begin{proof}
\textbf{Direction 1} ($\Leftarrow$): If $X_k$ is a co-tail of $X_i$, then the conditions hold.

Zero multivariate unique information: Since $X_k$ is not a parent, child, or co-head of $X_i$ (but rather a co-tail), by the previous theorem, we have $\Pi^{V'}_R(\{\{X_k\}\}; X_i) = 0$.

Positive synergistic information: Consider the directed hyperedge $h$ with $X_k, X_i \in T(h)$ and $X_j \in H(h)$. By the hypergraph collider amplification property:
\begin{equation}
    I(X_k; X_i \mid X_j) > I(X_k; X_i)
\end{equation}
Using the same PID decomposition argument as in Theorem \ref{thm:coparent_detection}, this inequality implies:
\begin{equation}
    \Pi^{\{X_j,X_k\}}_R(\{\{X_j, X_k\}\}; X_i) > 0
\end{equation}
Since $X_j$ is in the head of the hyperedge containing $X_i$ in its tail, $X_j$ is a child of $X_i$, ensuring $\Pi^{V'}_R(\{\{X_j\}\}; X_i) > 0$ by the previous theorem.

\textbf{Direction 2} ($\Rightarrow$): If the conditions hold, then $X_k$ is a co-tail of $X_i$.

We proceed by contradiction similar to Theorem \ref{thm:coparent_detection}. The key distinction is that the Markov blanket structure in hypergraphs differs from ordinary graphs. If $X_k$ is not a co-tail of $X_i$, then even if they share common head variables, they belong to separate hyperedges in the factorization. As a result, conditioning on the head part of the hyperedge induces conditional independence between $X_i$ and $X_k$ due to the global markov property.This conditional independence, combined with the information monotonicity property, leads to zero bivariate synergistic information, contradicting the assumed positive synergy. Therefore, $X_k$ must be a co-tail of $X_i$.
\end{proof}
\subsection{Proof of Theorem~\ref{thm:hyperedge_pid_signature}}\label{proof of theorem 6}

\begin{proof}
The theorem follows directly from our previous results:

\textbf{Property 1}: By Theorem \ref{thm:hypergraph_cotail_detection}, co-tail variables $X_k$ and $X_i$ exhibit positive bivariate synergistic information with their common head $X_j$.

\textbf{Property 2}: By Theorem \ref{thm:hypergraph_unique_info}, head variables $X_j$ are children of tail variables $X_i$, ensuring positive unique information.

\textbf{Property 3}: By Theorem \ref{thm:hypergraph_unique_info}, co-head variables are directly connected through their shared hyperedge membership, guaranteeing mutual unique information.
\end{proof}

\end{document}